\documentclass[twocolumn,aps,pre,groupedaddress,showpacs,showkeys]{revtex4}
\usepackage[T1]{fontenc}
\usepackage[latin1]{inputenc}
\usepackage{graphics}

\makeatletter

\providecommand{\LyX}{L\kern-.1667em\lower.25em\hbox{Y}\kern-.125emX\@}
\let\SF@@footnote\footnote
\def\footnote{\ifx\protect\@typeset@protect
    \expandafter\SF@@footnote
  \else
    \expandafter\SF@gobble@opt
  \fi
}
\expandafter\def\csname SF@gobble@opt \endcsname{\@ifnextchar[
  \SF@gobble@twobracket
  \@gobble
}
\edef\SF@gobble@opt{\noexpand\protect
  \expandafter\noexpand\csname SF@gobble@opt \endcsname}
\def\SF@gobble@twobracket[#1]#2{}

\usepackage[T1]{fontenc}
\usepackage[latin1]{inputenc}

\makeatletter

\usepackage{natbib}
\usepackage{graphicx}
\usepackage{bm}
\usepackage{amsfonts}
\usepackage{amssymb}
\usepackage{amsmath}
\def\dd{\! : \!}
\def\pd{\! \cdot \!}
\def\ta{{\bm \tau}}
\def\ep{{\bm \varepsilon}}
\def\ce{{\bf C}}
\def\iks{{\bm x}}
\def\ka{{\bm k}}
\def\eu{{\bf e}}
\def\uu{{\bf u}}
\def\G{{\bf G}}
\def\dc{\delta {\bf c}}
\def\aa{{\bf A}}
\def\bb{{\bf B}}
\def\Me{{\bf M}}
\def\te{{\bf T}}
\def\vv{{\bf W}}
\def\es{{\bf S}}
\makeatother
\makeatother
\makeatother
\makeatother
\makeatother
\makeatother

\makeatother
\makeatother
\makeatother

\makeatother

\begin{document}

\title{Fracture of disordered solids in compression as a critical phenomenon:
\\ II. Model Hamiltonian for a population of interacting cracks}

\author{Renaud Toussaint} \email[email: ]{Renaud.Toussaint@fys.uio.no} 
\altaffiliation[Present address: ]{ Department of Physics, University of Oslo, 
P.O. Box 1048 Blindern, 0316 Oslo 3, Norway} 
\affiliation{G\'eosciences Rennes, Universit\'e de Rennes 1, 35042 Rennes Cedex, France}

\author{Steven R. Pride} \email[email: ]{Steve.Pride@univ-rennes1.fr} 
\affiliation{G\'eosciences Rennes, Universit\'e de Rennes 1, 35042 Rennes Cedex, France}

\date{\today}

\begin{abstract} 

To obtain the probability distribution of 2D crack patterns in mesoscopic regions of a 
disordered solid, the formalism of Paper I requires that a functional form associating 
the crack patterns (or states) to their formation energy be developed. The crack states 
are here defined by an order parameter field representing both the presence and 
orientation of cracks at each site on a discrete square network. The associated 
Hamiltonian represents the total work required to lead an uncracked mesovolume  
into that state as averaged over the initial quenched disorder.  The effect of cracks 
is to create mesovolumes having internal heterogeneity in their elastic moduli.  
To model the Hamiltonian, the effective elastic moduli corresponding to a given 
crack distribution are determined that includes crack-to-crack interactions.     
The interaction terms  are entirely responsible for the localization transition 
analyzed in Paper III.  The crack-opening energies are related to these effective 
moduli via Griffith's criterion as  established in Paper I. 

\end{abstract} 


\maketitle

\section{INTRODUCTION\label{sec:intro} }

In triaxial-stress experiments on  rocks in the brittle regime, the onset
of a macroscopic localization of deformation is usually observed around peak
stress ({\em c.f}.\ B\'{e}suelle \cite{Bes99} for a review). Such departure from a
macroscopically uniform deformation regime is intrinsically beyond the capacities
of a mean-field theory, and so a specific model is developed here that takes
the orientational nature of crack-to-crack interactions into account. 

This is
the second paper in a series of three dedicated to exploring how the physical
properties of disordered solids evolve as they are led to failure in a state
of compression. The goal of this paper is to obtain a reasonable form for the
Hamiltonian \( E_{j}\left( \ep ,\ep_m\right)  \) which is defined as the
average work required to lead an intact region at zero deformation to the crack state
denoted by \( j \) 
when the maximum applied strain is $\ep_m$ and where the final strain $\ep$ is  
possibly  different than $\ep_m$  due to  a final unloading. 
This Hamiltonian must be expressed in terms of the spatial distribution of the
local order parameter that is the variable used to characterize the population
of cracks in each mesovolume of a huge disordered-solid system.  

Most existing lattice models explore the dynamics of scalar order parameters
either representing the breakdown of elastic spring or beam networks under tensile stress
\cite{AS93a}, 
 or of fuse networks \cite{BH98}. The analogies between such scalar models
and fracture of disordered media has been widely discussed \cite{HR90}. One 
advantage of our approach 
 is the ability to explore interactions based
on a fully tensorial description of the stress perturbations produced by each
crack.  Another is its ability to yield analytical rather than only numerical results. 
Using $E_j$ in the partition function established in
Paper I, it is possible to explore the crack patterns that emerge in compressive
settings for which isolated cracks appear in an intrinsically stable manner
no matter their size \cite{Loc95}, and for which macroscopic localization is
a collective phenomenon due to the energetic organization of small cracks as
opposed to an instability associated with the largest defects. In the present
paper, we  retain the leading-order effects of oriented-crack populations
interacting in 2D. The overriding importance of the long-range elastic interactions
leaves hope that 3D generalizations would not yield qualitatively different
critical behavior.


\section{PRINCIPLES OF THE MODEL\label{sec:principles}}

\subsection{Order-parameter definition\label{ssec:field_var}}

\begin{figure}
\resizebox*{7cm}{!}{\includegraphics{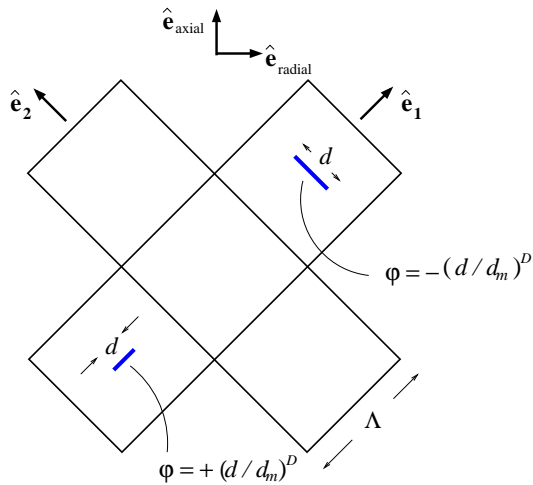} }

\caption{Part of the diamond network of cells that comprise a mesovolume. Each cell
has the linear dimension \protect\( \Lambda \protect \) and is only allowed
to contain one crack. The maximum length of any crack is \protect\( d_{m}\protect \)
and this length is assumed to be sufficiently small that \protect\( (d_{m}/\Lambda )^{D}\ll 1\protect \).
The amplitude of the order parameter is by definition \protect\( \psi =|\varphi |=(d/d_{m})^{D}\protect \)
where \protect\( d\protect \) is the length of the crack found in the cell,
while the sign of \protect\( \varphi \protect \) indicates the orientation
of the crack as shown. }

\label{orderparameter}
\end{figure}

We now elaborate on the  crack model introduced in Paper I.  
Each  mesovolume  of a huge rock system is discretized
into a square network of diamond-shaped cells of size \( \Lambda  \) (grain
sizes)  and only a single crack is potentially   present in each cell.
A  crack is located at the center of the cell
and has  a length \( d \) somewhere within the support \( [0,d_{m}] \)
where \( d_{m} \) is the maximum crack length (a fixed parameter of the system)
required to satisfy \( d_{m}<\Lambda  \). In the perturbative treatment of
the crack interactions developed herein,  \( \epsilon =(d_{m}/\Lambda )^{D} \)
is taken to be a small number where \( D \) is the number of space-dimensions (in the present
model, \( D=2 \)). The local order parameter \( \varphi({\bm x})  \) associated with each cell  
${\bm x}$ is taken to have an amplitude \( \psi =|\varphi |=(d/d_{m})^{D} \) and has a  
 sign that  indicates whether the crack is oriented
at \( +45^{\circ } \) or \( -45^{\circ } \) relative to the principle-stress direction 
 (the so-called ``axial direction'').  
The  model is summarized in Fig.\ \ref{orderparameter}

The restriction that cracks are either at \( \pm 45^{\circ } \) and have lengths
less than the grain dimensions is of course a great simplification compared
to what is found inside of real rocks. However, we only need  to characterize
 the essential features of a crack population that contribute to localization
phenomena and, to this extent, it appears overly complicated to  model  
the amazing variety of crack geometries encountered in real rocks \cite{Kra83,BSS80}.
The localization  transition involves spontaneous breaking of symmetries
both under translation and parity (inversion of the minor stress axis) as is
seen from the structure of the shear bands emerging in the post-peak-stress
regime \cite{Pat78}. The essential feature of any proposed order parameter
is that it must reflect and quantify the amount of local symmetry breaking and
our simple model with cracks at either  \( \pm 45^{\circ } \) does just
this.

Furthermore, there is  evidence
both from acoustic-emissions monitoring \cite{LBK+92} and from direct observation
after unloading \cite{MME00}, that cracks developing prior to peak stress do
not exceed an extent of a few grains diameters.  
 This is principally because the grain contacts which break  are
much weaker than the grains and have a finite length so that cracks  arriving  
in compression do so stably \cite{Loc95}.   Crack coalescence is  not 
explicitly allowed for.  However, since several neighboring cells in a line may all 
contain cracks of the same sign, the long-range elastic effect of 
 long (coalesced) cracks is effectively allowed for.  
 Our picture of the final shear bands
 experimentally observed in the post-peak-stress regime is that they were created
by unstable sliding  along a band weakened  in the pre-peak-stress
regime by a concentration of  coherently-oriented cracks \cite{Pat78}. Our
model allows  small cracks to stably concentrate \emph{en \'{e}chelon}
along conjugate bands relative to the principal-stress direction;  
  however, it does not model  the final  
 unstable rupture along  a given  band.

\subsection{Formation energy of a crack pattern\label{ssec:Form} }

It has been established in Paper I that to a reasonable approximation, the work
required to form a crack state, as averaged over the initial disorder, separates
into one part representing the work required to break the grain contacts, and
a second part representing the elastic energy stored in the cracked solid. This
was expressed in Eqs.\ (26)--(28) of Paper I as  
\begin{equation}
\label{hamiltonian}
E_{j}=\frac{q}{2}{\bm \varepsilon_{m}}\dd \left( {\textbf {C}}_{0}-{\textbf {C}}_{j}\right) 
\dd {\bm \varepsilon_{m}}+\frac{1}{2}{\bm \varepsilon }\dd {\textbf {C}}_{j}\dd {\bm \varepsilon }.
\end{equation}
 The first term of Eq.\ (\ref{hamiltonian}) is the energy spent in the irreversible
formation of the crack state \( j \) averaged over quenched disorder and was obtained 
through an application of  
 Griffith's principle. The parameter $q$ derives from the  quenched-disorder 
distribution and lies in the range $[1/2,1]$ (see Sec.\ III B.2 of Paper I). 
The second term  
is the reversibly-stored elastic energy with \( {\textbf {C}}_{j} \) being
the elastic-stiffness tensor of state \( j \).

Our principal task is therefore to model the way that cracks and collective
crack states affect the overall elastic moduli of a mesovolume. This requires
detailed knowledge of the stress (or strain) field throughout the mesovolume
in the presence of arbitrary crack populations, and we treat this need using
the following approximations. First, since the cracks in the model are isolated
one to each cell, their main effect regarding the far-field stress is to change
the elastic moduli of their embedding cell. Such a change is modeled assuming
the cracks to be penny-shaped ellipsoidal cavities. 
 We ignore  how such ellipses change 
shape when the  applied stress is unloaded/reloaded  
 since linear elasticity  alone captures 
the principle effect of how  the rock  becomes weaker due to 
strategic placement of cracks in cells.   
Since a crack  occupies a limited extent of a cell, the modification
of the moduli is small compared to the moduli of the intact cell so that the
resulting far-field stress field can be developed as a Born series. It is in
the third term of this development that crack-to-crack interactions are  first
allowed for. Higher-order  interactions (three cracks simultaneously interacting
and so on) are negligible to the extent that \( \epsilon =(d_{m}/\Lambda )^{D} \)
can be considered  small.

\section{ELASTIC ENERGY\label{sec:Elast_En}}

The goal of this section is to determine the elastic energy \( E^{\text {el}}_{j} \)
stored in a mesovolume occuping the region \( \Omega  \) and containing the
crack state \( j \) (which denotes the spatial distribution of \( \varphi (\iks ) \)
at all points \( \iks  \) of \( \Omega  \)) when a displacement corresponding
to a uniform strain tensor \( \ep ^{(0)} \) is applied on the external surface
\( \partial \Omega  \) of the mesovolume.

\subsection{Elastic energy of a weakly heterogeneous  solid \label{sssec:en,from,dc}}

The effect of the crack field \( \varphi ({\iks }) \) is to perturb the stiffness
tensor of each cell as \( \ce (\iks )=\ce ^{0}+\dc \left[ \varphi (\iks )\right]  \)
where \( \ce ^{0} \) denotes the moduli of an uncracked cell (assumed uniform
for all cells), and where \( \dc (\iks ) \) is a small perturbation due to
the possible presence of a crack as characterized by \( \varphi (\iks ) \).
It is established in Appendix \ref{ssec:Modulii} that the non-zero components
of \( \dc  \) are typically smaller than those of \( \ce ^{0} \) by a factor
\( \epsilon =\left( d_{m}/\Lambda \right) ^{D}\ll 1 \). Our problem is to resolve
an elasticity boundary-value problem in a region \( \Omega  \) containing a
weakly heterogeneous stiffness tensor \( \ce ({\iks }) \).

The displacement boundary conditions are given as 
\begin{equation}
\label{eq:bond,cond,displ}
\forall \iks \in \partial \Omega ,\, \, \, \, \uu (\iks )=\ep ^{(0)}\pd {\iks }
\end{equation}
 where \( \iks  \) denotes distance from the center of the mesovolume. Elastostatic
equilibrium requires that 
\begin{equation}
\label{eq:elastostatic,eq}
\partial _{j}\tau _{ij}=C_{ijkl}^{0}\partial _{j}\partial _{k}u_{l}+\partial _{j}[\delta c_{ijkl}\partial _{k}u_{l}]=0
\end{equation}
 throughout \( \Omega  \) where summation over repeated indices is assumed
both here and throughout. Due to the linearity of the problem, we use the elastostatic
Green tensor \( \G =G_{ij}\hat{\iks }_{i}\hat{\iks }_{j} \) for a uniform material
which is a solution of 
\begin{eqnarray}
C_{ijkl}^{0}\partial _{j}\partial _{k}G_{lm}(\iks ,\iks ')+\delta _{im}\delta ^{D}(\iks -\iks ') & = & 0\label{eq:def,Green} \\
\forall \iks \in \partial \Omega ,\, \, \, \, G_{ij}(\iks ,\iks ') & = & 0.\label{eq:bound,cond,Green} 
\end{eqnarray}
 The components \( G_{ij}(\iks ,\iks' ) \) define the \( i^{\textrm{th}} \)
component of the displacement at \( \iks  \) induced by a unit point force
acting along the \( j \)-axis at \( \iks ' \). Here, \( \delta _{ij} \) is
the Kronecker symbol, and \( \delta ^{D} \) is the \( D \)-dimensional Dirac
distribution.

The solution for the displacements when no cracks are present is simply \( \uu ^{(0)}(\iks )=\ep ^{(0)}\pd \iks  \)
throughout all of \( \Omega  \). Thus, it is a straightforward excercise to
demonstrate that the total displacement \( \uu  \) in the presence of the cracks
satisfies the following integral equation 
\begin{equation}
\label{eq:rel,u,green}
u_{i}(\iks )=u^{(0)}_{i}(\iks )+\! \! \int _{\Omega }\! \! G_{ij}(\iks ,\iks ')\partial _{k'}[\delta c_{jklm}\partial _{l'}u_{m}](\iks ')d^{D}\iks '
\end{equation}
 where \( \partial _{i'} \) denotes the partial derivative relative to the
coordinate \( x'_{i} \). Using \( \epsilon  \) as the argument of a series
expansion, we write the displacements as \( \uu =\uu ^{(0)}+\uu ^{(1)}+\cdots \uu ^{(n)}+O(\epsilon ^{(n+1)}), \)
where each \( \uu ^{(m)} \) is \( O(\epsilon ^{m}) \). Collecting terms at
each order of \( \epsilon  \) in Eq.\ (\ref{eq:rel,u,green}) gives the following
recursion relation 
\begin{equation}
\label{eq:recurrence,pour,u,n}
u^{(n+1)}_{i}(\iks )=\int _{\Omega }G_{ij}(\iks ,\iks ')\partial _{k'}[\delta c_{jklm}\partial _{l'}u^{(n)}_{m}](\iks ')d^{D}\iks '.
\end{equation}
 The boundaries conditions used to define \( \G  \) guarantee that for all
\( n>0 \), the displacements \( \uu ^{(n)} \) are zero on the boundary \( \partial \Omega  \).

The quantity we are specifically seeking to establish is the elastic energy
density \( E^{\textrm{el}}={\ell ^{-D}}\int _{\Omega }\frac{1}{2}\ta (\iks )\dd \ep (\iks )d^{D}\iks  \)
where we recall that \( \ell  \) is the linear dimension of a mesovolume. The
definitions of the strain \( \ep _{ij}=\frac{1}{2}(\partial _{i}\uu _{j}+\partial _{j}\uu _{i}) \)
and stress \( \tau _{ij}=(C_{ijkl}^{0}+\delta c_{ijkl})\varepsilon _{kl} \)
give immediately the following relations: 
\begin{eqnarray}
\varepsilon ^{(n)}_{ij} & = & \frac{1}{2}(\partial _{i}u^{(n)}_{j}+\partial _{j}u^{(n)}_{i})\label{eq:terms,strain} \\
\tau ^{(n)}_{ij} & = & C_{ijkl}^{0}\varepsilon ^{(n)}_{kl}+\delta c_{ijkl}\varepsilon _{kl}^{(n-1)}\label{eq:terms,stress} \\
E^{(n)} & = & \frac{1}{2\ell ^{D}}\sum ^{n}_{a=0}\int _{\Omega }\ta ^{(n-a)}\dd \ep ^{(a)}d^{D}\iks \label{eq:terms,E} 
\end{eqnarray}
 with the convention \( \ep ^{(-1)}=0 \). In the last expression, the fact
that \( \uu ^{(a)}=0 \) on the boundary for all \( a>0 \) guarantees that after integrating 
by parts
\begin{eqnarray*}
\int _{\Omega }\ta ^{(n-a)}\dd \ep ^{(a)}d^{D}\iks & = & \int _{\Omega }\ta ^{(n-a)}\dd \nabla \uu ^{(a)}d^{D}\iks \\
 & = & \int _{\partial \Omega }{\textbf {n}}\pd {\ta }^{(n-a)}\pd {\uu }^{(a)}d^{D-1}\iks = 0 
\end{eqnarray*}
 where we used the facts that the stress tensor is symmetric and solenoidal.
The \( n^{\textrm{th}} \) term of the total elastic energy is then 
\begin{equation}
\label{eq:En}
E^{(n)}=\frac{1}{2}\, \overline{\ta ^{(n)}}\dd \ep ^{(0)}
\end{equation}
 where the upper bar denotes a volume average over \( \Omega  \).

The first term of the elastic energy is independent of the \( \varphi  \) field,
and corresponds to the physically unimportant amount of energy 
\begin{equation}
\label{eq:E0}
E^{(0)}=\frac{1}{2}\ep ^{(0)}\dd \ce ^{(0)}\dd \ep ^{(0)}
\end{equation}
 stored in the intact state.

For the higher orders \( n\geq 1 \) , \( \ta ^{(n)} \) is expressed by Eq.\  (\ref{eq:terms,stress}),
and the same argument as above using the fact that \( \uu ^{(n)}=0 \) on \( \partial \Omega  \)
eliminates a term: 
\begin{eqnarray}
E^{(n)} & = & \overline{\ep ^{(n)}}\dd \ce ^{(0)}\dd \ep ^{(0)}+\frac{1}{2}\overline{\ep ^{(n-1)}\dd \dc }\dd \ep ^{(0)}\nonumber \\
 & = & \frac{1}{2}\overline{\ep ^{(n-1)}\dd \dc }\dd \ep ^{(0)}.\label{eq:En.gt.1} 
\end{eqnarray}
 The second term of the developement, 
\begin{equation}
\label{eq:E1}
E^{(1)}=\frac{1}{2}\ep ^{(0)}\dd \overline{\dc }\dd \ep ^{(0)}
\end{equation}
 represents only a local dependance on \( \dc  \) (and therefore on the crack-field) 
since  it does not involve nested integrals over two different
positions. It will be shown to represent  only the contribution of the average
crack porosity to the stiffness of the rock.

The third term of the development is where the desired crack-to-crack interactions
arrive. Using the symmetry of \( \dc  \) under the inversion of its two first
or last indices, and Eq.\ (\ref{eq:recurrence,pour,u,n}) to have an integral
form of \( \uu ^{(1)} \), Eq.\ (\ref{eq:En.gt.1}) transforms to 
\begin{eqnarray}
2\ell ^{D}E^{(2)} & = & \! \! \int \! \partial _{b}u^{(1)}_{a}(\iks )\delta c_{abcd}(\iks )\varepsilon _{cd}^{(0)}d^{D}\iks \nonumber \\
 & = & \! \! \int \! \! \! \int \! \varepsilon _{cd}^{(0)}\delta c_{abcd}(\iks )\partial _{b}G_{aj}(\iks ,\iks' )\nonumber \\
 &  & \quad \times \partial _{k'}\delta c_{jklm}(\iks ')\varepsilon _{lm}^{(0)}\, 
d^{D}\iks \, d^{D}\iks'.\label{eq:E2} 
\end{eqnarray}
 This term accounts for the way that a crack present at \( \iks ' \) energetically
interacts with a different crack at \( \iks  \). This is the non-local interaction
term that is ultimately responsible for the localization transition. The higher
terms of the Born development can be neglected for our purposes.

\subsection{Elastic energy as explicit function of the crack field\label{sssec:en,with,strain}}

To establish the terms of the Born-approximated elastic energy as explicit functions
of both the crack state \( \varphi  \) and the imposed strain \( \ep ^{(0)} \),
a few definitions are first introduced.

The principal axes of \( \ep ^{(0)} \) are along \( \left( \hat{\eu }_{\text {radial}},\hat{\eu }_{\text {axial}}\right)  \)
as denoted in Fig.\ \ref{orderparameter}. Our square network of cells is rotated
\( +45^{\circ } \) from this orthonormal basis. We work here in the coordinates
\( \left( \hat{\eu }_{1},\hat{\eu }_{2}\right)  \) of the square network so
that the applied strain takes the form 
\begin{equation}
\label{eq:applied,eps}
\ep ^{(0)}=\frac{1}{2}\left( \begin{array}{cc}
\Delta  & \gamma \\
\gamma  & \Delta 
\end{array}\right) 
\end{equation}
 where \( \Delta =\varepsilon _{\text {radial}}+\varepsilon _{\text {axial}} \)
and \( \gamma =-(\varepsilon _{\text {radial}}-\varepsilon _{\text {axial}}) \)
are the imposed dilatation and shear strain.

For convenience, we assume the intact material to be isotropic. Taking \( \lambda +2\mu  \)
as the stress unit, where \( (\lambda ,\mu ) \) are the Lam\'{e} parameters
of the material, and using the usual tensor-to-matrix mapping of the indices
\( (11)\rightarrow 1 \); \( (22)\rightarrow 2 \); \( (12)\rightarrow 3 \),
the fourth-order stiffness tensor of the intact material takes the form 
\begin{equation}
\label{eq:C0}
\ce ^{0}=\left( \begin{array}{ccc}
1 & 2\alpha -1 & 0\\
2\alpha -1 & 1 & 0\\
0 & 0 & 1-\alpha 
\end{array}\right) 
\end{equation}
 where 
\begin{equation}
\label{eq:def,alpha}
\alpha =\frac{\lambda +\mu }{\lambda +2\mu }
\end{equation}
 is a material-dependent constant in the range \( [0.5,1] \).

The deviation \( \dc  \) of this tensor due to the possible presence of a crack
in a cell separates into an isotropic contribution independant of the crack's
orientation, and into an anisotropic orientation-dependent contribution. In
Appendix \ref{ssec:Modulii}, we demonstrate that 
\begin{eqnarray}
\lefteqn {\dc (\iks )=\epsilon \left[ \aa \varphi (\iks )+\bb |\varphi (\iks )|\right] } &  & \label{eq:C,A,B,phi} \\
\aa  & \! \! = & \! \! \! \left( \begin{array}{ccc}
\eta _{2}-\eta _{1} & 0 & 0\\
0 & \eta _{1}-\eta _{2} & 0\\
0 & 0 & 0
\end{array}\right) \label{eq:A,simpl} \\
\bb  & \! \! = & \! \! \left( \begin{array}{ccc}
-\eta _{1} & -(2\alpha -1)\eta _{2} & 0\\
-(2\alpha -1)\eta _{2} & -\eta _{1} & 0\\
0 & 0 & -(1-\alpha )\eta _{3}
\end{array}\right) \label{eq:B,simpl} 
\end{eqnarray}
 where \( (\eta _{1},\eta _{2},\eta _{3}) \) are positive constants expressed
in Appendix \ref{ssec:Modulii} in terms of the Lam\'{e} parameters.

Making the necessary contractions over the indices, we easily obtain the trivial
(crack-independent) energy \( E^{(0)} \) using Eqs.\ (\ref{eq:E0}), (\ref{eq:applied,eps})
and (\ref{eq:C0}). For later convenience, this result is best written in matrix
form as 
\begin{eqnarray}
E^{(0)} & = & \frac{1}{2}(\Delta ,\gamma )\pd \Me _{0}\pd (\Delta ,\gamma )^{T}\label{eq:E0,delta,gamma} \\
\Me _{0} & = & \left( \begin{array}{cc}
\alpha  & 0\\
0 & 1-\alpha 
\end{array}\right) .\label{eq:M0} 
\end{eqnarray}
 Using the auxiliary field 
\begin{equation}
\label{eq:def,psi,from,phi}
\psi (x)=\left| \varphi (x)\right| 
\end{equation}
 denoting the amplitude of each crack, one similarly obtains {[}using Eqs.\  (\ref{eq:E1})
and (\ref{eq:C,A,B,phi})-(\ref{eq:B,simpl}){]} 
\begin{eqnarray}
E^{(1)} & = & \frac{\epsilon }{2}\left[ \ep ^{(0)}\dd \aa \dd \ep ^{(0)}\, \overline{\varphi }+\ep ^{(0)}\dd \bb \dd \ep ^{(0)}\, \overline{\psi }\right] \nonumber \\
 & = & \frac{1}{2}(\Delta ,\gamma )\pd \Me _{1}\pd (\Delta ,\gamma )^{T}\label{eq:E1,delta,gamma} 
\end{eqnarray}
 with 
\begin{eqnarray}
\Me _{1} & = & -\epsilon \overline{\psi }\left( \begin{array}{cc}
\kappa _{2} & 0\\
0 & \kappa _{3}
\end{array}\right) \label{eq:M1} \\
\kappa _{2} & = & \frac{\eta _{1}}{2}+\frac{2\alpha -1}{2}\eta _{2}\label{eq:kappa2} \\
\kappa _{3} & = & (1-\alpha )\eta _{3}.\label{eq:kappa3} 
\end{eqnarray}
 The term proportional to \( \overline{\varphi } \) has algebraically cancelled
due to the symmetry of the problem under parity; inversion of the minor axis
\( \widehat{\textbf {e}}_{\text {radial}} \)flips the orientation of cracks,
and therefore changes the sign of \( \overline{\varphi } \), while the energy
remains necessarily unchanged. The surviving term is negative and proportional
to \( \overline{\psi } \), and accounts for the softening of the mesovolume
due to the presence of cracks. This dependence on the total number of cracks
is the only order-parameter dependent effect to first order.

Last, the crack-interaction term of principal interest can be readily expressed
from Eqs.\  (\ref{eq:E2}) and (\ref{eq:C,A,B,phi}) as 
\begin{eqnarray}
-2\ell ^{D}E^{(2)} & = & \epsilon ^{2}\ep _{cd}^{(0)}\varepsilon _{kl}^{(0)}A_{abcd}A_{ijkl}f_{aibj}\nonumber \\
 &  & +2\epsilon ^{2}\varepsilon _{cd}^{(0)}\varepsilon _{kl}^{(0)}A_{abcd}B_{ijkl}g_{aibj}\nonumber \\
 &  & +\epsilon ^{2}\varepsilon _{cd}^{(0)}\varepsilon _{kl}^{(0)}A_{abcd}B_{ijkl}h_{aibj}\label{eq:E2,bis} 
\end{eqnarray}
 where the fourth-order tensors \( {\textbf {f}},{\textbf {g}},{\textbf {h}} \)
are functionals of \( \varphi  \) and defined as 
\begin{eqnarray}
f_{aibj} & = & \int d^{D}\iks \int d^{D}\iks '\, G_{ai}(\iks ,\iks ')\partial _{b}\varphi \partial _{j'}\varphi \label{eq:aux1} \\
g_{aibj} & = & \int d^{D}\iks \int d^{D}\iks '\, G_{\{ai\}}(\iks ,\iks ')\partial _{b}\varphi \partial _{j'}\psi \label{eq:aux2} \\
h_{aibj} & = & \int d^{D}\iks \int d^{D}\iks '\, G_{ai}(\iks ,\iks ')\partial _{b}\psi \partial _{j'}\psi .\label{eq:aux3} 
\end{eqnarray}
 In the second term, the reciprocity of the Green function \( G_{ai}(\iks ,\iks ')=G_{ai}(\iks ',\iks ) \)
is used as well as the notation \( G_{\{ai\}}=(G_{ai}+G_{ia})/2 \).

The Green tensor needed here satisfies the Dirichlet conditions of Eq.\ (\ref{eq:bound,cond,Green})
and can be obtained, in principle, from the infinite-space Green tensor via
the image method. However, this transforms \( E^{(2)} \) into an infinite series
(one term for each image), and makes the functional integrations of Paper III
analytically hopeless. To remedy this problem, the Green function with a periodic
instead of zero boundary condition is used as ersatz. Since \( \uu ^{(1)} \)
is only affected close to the boundaries by this replacement, this approximation
will be considered valid for the evaluation of the volume integral \( E^{(2)} \).

The double integrals of Eqs.\ (\ref{eq:aux1})--(\ref{eq:aux3}) are most easily
expressed using the 2D finite-Fourier transform 
\begin{eqnarray}
\widetilde{F}(\ka ) & = & \int _{\Omega }d^{D}\iks \, F(\iks )e^{-i\ka .\iks }\label{eq:Tfourier} \\
F(\iks ) & = & \frac{1}{\ell ^{D}}\sum _{\ka }\widetilde{F}(\ka )e^{i\ka .\iks }\label{eq:Tfourierreciproqie} 
\end{eqnarray}
 where the sum over the wavevectors \( \ka  \) is over \( \{\ka ={2\pi }n_{i}/\ell \hat{\eu }_{i}\, ;\, \forall i,\, n_{i}\in \mathbb {Z}\} \)
with an upper cutoff given by \( \max n_{i}>\ell /\Lambda  \) that reflects
the fact that the order-parameter cannot vary on scales smaller than cell sizes
\( \Lambda  \). Since the Green function used is defined with periodic boundary
conditions, it satisfies \( G_{ai}(\iks ,\iks ')=G_{ai}(\iks -\iks ') \). Its
Fourier transform is easily established,  
and upon recalling
that \( (\lambda +2\mu ) \) is adopted as the stress unit, reads 
\begin{eqnarray}
\widetilde{\G }(\ka ) & = & \frac{1}{(1-\alpha )\ka ^{2}}({\textbf {I}}-\alpha \hat{\ka }\hat{\ka })\label{eq:green2D,bis} \\
\hat{\ka } & = & \frac{\ka }{\left\Vert \ka \right\Vert }\label{eq:dirk} 
\end{eqnarray}
 where \( {\textbf {I}} \) is the identity tensor. This is real and symmetric,
as is \( \G (\iks ) \) itself. Since \( \varphi  \) and \( \psi  \) are real
fields, one has \( \widetilde{\varphi }(-\ka )=\widetilde{\varphi }^{*}(\ka )\, \mbox {and}\, \widetilde{\psi }(-\ka )=\widetilde{\psi }^{*}(\ka ) \).
Using these relations, together with the identity \( \int _{\Omega }d^{D}\iks \, e^{i\ka .\iks }=\ell ^{D}\delta _{\ka } \),
the integrals of Eqs.\ (\ref{eq:aux1})--(\ref{eq:aux3}) become the following
sums 
\begin{eqnarray}
f_{aibj} & = & \frac{1}{\ell ^{D}}\sum _{\ka \neq 0}|\widetilde{\varphi }(\ka )|^{2}\hat{\ka }_{j}\hat{\ka }_{b}(\delta _{ia}-\alpha \hat{\ka }_{i}\hat{\ka }_{a})\label{eq:aux1bis} \\
g_{aibj} & = & \frac{1}{\ell ^{D}}\sum _{\ka \neq 0}\Re [\widetilde{\varphi }(\ka )
\widetilde{\psi }^{*}(\ka )]\hat{\ka }_{j}\hat{\ka }_{b}
(\delta _{ia}-\alpha \hat{\ka }_{i}\hat{\ka }_{a})\label{eq:aux2bis} \\
h_{aibj} & = & \frac{1}{\ell ^{D}}\sum _{\ka \neq 0}|\widetilde{\psi }(\ka )|^{2}\hat{\ka }_{j}\hat{\ka }_{b}(\delta _{ia}-\alpha \hat{\ka }_{i}\hat{\ka }_{a})\label{eq:aux3bis} 
\end{eqnarray}
 where \( \Re  \) denotes the real part of a complex quantity and where \( \widehat{\textbf {k}}_{i} \)
denotes the \( i^{\textrm{th}} \) component of \( \widehat{\textbf {k}}=\ka /\Vert \ka \Vert  \).
With the following definitions associated with the orientation of \( \ka  \)
\begin{eqnarray}
\theta _{\ka } & = & (\widehat{{\bm e}_{1},\ka })\label{eq:def,thetak} \\
u_{\ka } & = & \cos (2\theta _{\ka })=\cos ^{2}\theta _{\ka }-\sin ^{2}\theta _{\ka }=\hat{\ka }^{2}_{1}-\hat{\ka }^{2}_{2}\label{eq:def,uk} \\
v_{\ka } & = & \sin (2\theta _{\ka })=2\cos \theta _{\ka }\sin \theta _{\ka }=2\hat{\ka }_{1}\hat{\ka }_{2}\label{eq:def,vk} 
\end{eqnarray}
 the remaining contraction in Eq.\  (\ref{eq:E2,bis}) over the eight indices
\( (abcdijkl) \) is performed. The calculation is a bit long but without surprise
and finally produces 
\begin{eqnarray}
E^{(2)} & = & \frac{1}{2}(\Delta ,\gamma )\pd \Me _{2}\pd (\Delta ,\gamma )^{T}\label{eq:E2,delta,gamma} \\
\Me _{2} & = & \frac{-\epsilon ^{2}}{(1-\alpha )\ell ^{2D}}\left( \begin{array}{cc}
a & b\\
b & c
\end{array}\right) \label{eq:M2} 
\end{eqnarray}
 where the components \( a \), \( b \), and \( c \) are defined \begin{gather}
 a = \sum _{\ka \neq 0} a_{\ka} \,  ; \quad 
\quad b = \sum _{\ka \neq 0} b_{\ka} \, ; \quad 
 \quad c = \sum _{\ka \neq 0} c_{\ka} \label{eq:abc}\\
\begin{split}
a_{\ka} = &
 \quad (1-\alpha u_{\ka }^{2})\kappa ^{2}_{1}\, |\widetilde{\varphi }_{\ka }|^{2}\\
& +2(1-\alpha )u_{\ka}\kappa _{1}\kappa _{2}\, 
\Re \!\left( \widetilde{\varphi }_{\ka }\widetilde{\psi }^{*}_{\ka }\right) \\
& +(1-\alpha )\kappa ^{2}_{2}\, |\widetilde{\psi }_{\ka }|^{2}  \\
b_{\ka} = &
-\alpha u_{\ka}v_{\ka}\kappa _{1}\kappa _{3} \, 
\Re\! \left( \widetilde{\varphi }_{\ka}\widetilde{\psi }^{*}_{\ka}\right) \\
& +(1-\alpha )v_{\ka}\kappa _{2}\kappa _{3}\, |\widetilde{\psi }_{\ka}|^{2} \\
c_{\ka} = &  (1-\alpha v^{2}_{\ka})\kappa _{3}^{2}\, |\widetilde{\psi }_{\ka}|^{2}  
\end{split}
\nonumber
\end{gather} with \( \kappa _{2},\kappa _{3} \) defined in Eqs.\  (\ref{eq:kappa2}),
(\ref{eq:kappa3}) and \( \kappa _{1} \) a new material-dependent constant,
\begin{equation}
\label{eq:kappa1}
\kappa _{1}\hat{=}\frac{\eta _{1}-\eta _{2}}{2}.
\end{equation}

\section{SURFACE FORMATION ENERGY\label{sec:Surf_En}}

Next, we must account for the energy \( E_{j}^{{I}} \) that irreversibly went
into  creating the cracks of a given crack state \( j \) at a maximum deformation
\( {\bm \varepsilon_m } \). In Paper I, this contribution was obtained using 
 Griffith's  criterion as  
\begin{equation}
\label{grif}
E_{j}^{{I}}= \frac{q}{2}{\ell ^{D}}
{\bm \varepsilon_m }\dd \left( {\textbf {C}}_{0}-{\textbf {C}}_{j}\right) \dd {\bm \varepsilon_m}
\end{equation}
 where $q$ derives from the quenched disorder and is  bounded as \( 0.5 \le q<1 \).
The derivation of this  statement implicitly assumed that all cracks were the same length.  
In the present treatment,  cracks are allowed to have any length in the range $0\le d \le d_m$.  
It is a straightforward excercise to demonstrate that if the breaking energies  
for each possible length $d$   are all sampled from the same quenched-disorder distribution, 
then Eq.\ (\ref{grif}) again holds.   We forego such a demonstration.   
In the notation of the present paper we may thus state that  
\begin{equation}
E[\varphi ]^{I}  =  q(E^{(0)}-E^{\textrm{el}}[\varphi ])
 =  -q(E^{(1)}+E^{(2)})[\varphi ]
\end{equation}
 where $E^{(1)}$ and $ E^{(2)}$ are the terms of the Born-development given  
by  Eqs.\ (\ref{eq:E1,delta,gamma}) and (\ref{eq:E2,delta,gamma})  
upon replacing the current  strain parameters $\Delta$ and $\gamma$ 
 by the maximum-achieved strain  
$\Delta_m$ and $\gamma_m$.

\section{Temperature\label{sec:temperature}}
 Although  not required as part of the model Hamiltonian, we now give 
an explicit in $\Delta_m, \gamma_m$ approximate 
expression for the temperature  by using  
 Eq.\ (59) of Paper I.  This temperature was derived in Paper I assuming    
only a single crack size.  Unfortunately, the result does not easily 
generalize to multiple crack sizes  and so  we simply take $d=d_m$ to obtain  
 the estimate 
\begin{equation}
\ell ^{D}T(\Delta_m,\gamma_m)=-\frac{ (1-q)
 d_m^D \, {e}_{1}(\Delta_m,\gamma_m) }
{\ln \left\{ [\zeta /{e}_{1}(\Delta_m,\gamma_m)]^{q/(1-q)}-1\right\} }. 
\end{equation}
where  
\( {e}_{1}d_{m}^{D} \) is how much the first-Born elastic energy in a mesovolume 
is reduced when a crack of length $d_m$ is introduced  
[{\em c.f.}\ Eqs.\ (\ref{eq:E1,delta,gamma}) and (\ref{eq:M1})].  The  energy density     
\( e_{1}(\Delta_m,\gamma_m) \) is defined 
\begin{equation}
{e}_{1}(\Delta_m,\gamma_m)=\frac{1}{2}\left(\kappa_2 \Delta_m^{2}+  
\kappa_3 \gamma_m^{2} \right) 
\end{equation}
 while \( \zeta  \) is a dimensionless ``fracture toughness'' parameter defined
as 
\begin{equation}
\zeta \equiv \frac{\Gamma }{ (\lambda + 2 \mu) d_{m}}.
\end{equation}
   There is a phase transition 
when \( (\zeta /e_1)^{q/(1-q)}=2 \)  and  $T$   
 diverges so that all crack states become equally probable.  

We now consider  whether such a phase transition is expected in 
laboratory experiments on rocks.   The 
order of magnitude values $\Gamma \sim 10^2$ J/m$^2$, $d_m\sim 10^{-5}$ m,  
$\kappa_2\sim 1$, and $\lambda+2\mu \sim 10^{10}$ Pa are appropriate for typical 
grains in rocks so that $\zeta \sim 10^{-3}$.          
When a rock fails in shear, the accumulated strain is  on the order 
of a percent or two so that   
 the maximum value of $e_1$ of interest is also on the order of $10^{-4}$.  We thus  find 
that at   shear failure,   
 \( \zeta /e_1 \sim 10\) and so  we {\em a priori} expect the localization 
transition to occur prior to the  
 temperature-divergence transition.  This is more quantitatively demonstrated in Paper III.

\section{SUMMARY\label{sec:summary}}

Collecting together both the elastic energy and the surface formation energy, 
we obtain at last the Hamiltonian to be
used in performing ensemble averages over crack states in the next paper. We
write this Hamiltonian in the final form 
\begin{eqnarray*}
E_{j}(\ep ,  \ep_m) & = & E^{R}(\ep )[\varphi ]+E^{I}(\ep_m)[\varphi ]\label{final,hamiltonian} \\
E^{R}(\ep )[\varphi ] & = & E^{0}(\Delta ,\gamma )+E^{\text {av}}(\Delta ,\gamma )[\varphi ]+E^{\text {int}}(\Delta ,\gamma )[\varphi ]\label{el,energy} \\
E^{I}(\ep_m)[\varphi ] & = & 
-q \left\{ E^{\text {av}}(\Delta_m,\gamma_m)[\varphi ]
+E^{\text {int}}(\Delta_m,\gamma_m)[\varphi ]\right\} \label{crack,energy} 
\end{eqnarray*}
where \( (\Delta ,\gamma ) \) are the isotropic and shear strain components of  the current  
strain tensor \( \ep  \), and \( (\Delta_m,\gamma_m) \) are similar quantities
referring to the maximum achieved strain \( \ep_m \). 
The energy \( E^{0} \) is the trivial elastic energy of the uncracked state
\begin{equation*}
\label{triv,energy}
E^{0}(\Delta ,\gamma )=\frac{1}{2}\{\alpha \Delta ^{2}+(1-\alpha )\gamma ^{2}\}
\end{equation*}
where  \( \alpha  \) is a dimensionless elastic constant in the range \( [0.5,1] \)
 defined by  Eq.\ (\ref{eq:def,alpha}).

The next term in the Born development is  
\( E^{\text {av}}=E^{(1)} \)  
which depends only on the volume average \( \bar{\psi } \), which is 
the  fraction of cracked
cells in the crack state \( \varphi  \) and  is thus entirely independent of
the spatial fluctuations of \( \varphi  \). Its dependence on the strain  
\( (\Delta ,\gamma ) \) is   
$$
E^{\text {av}}(\Delta ,\gamma )[\varphi ]
=-\frac{1}{2}[\kappa _{2}\Delta ^{2}+\kappa _{3}\gamma ^{2}]\epsilon \bar{\psi }.  
$$
 We defined \( \epsilon =(d_{m}/\Lambda )^{D} \) to be a small parameter where
\( D=2 \) is the number of space dimensions in the model, and \( d_{m} \),
\( \Lambda  \), and \( \ell  \) as respectively the linear sizes of the largest
crack, a unit cell, and a mesovolume. The three coefficients \( \kappa _{i} \)
are positive dimensionless material-dependent constants defined by Eqs.\ (\ref{eq:kappa2}),
(\ref{eq:kappa3}) and (\ref{eq:kappa1}).

The interaction energy  \(  E^{\text {int}} = E^{(2)}\) involve a quadratic
matrix operator \( P_{\ka } \) that, for each non-zero wave-vector \( \ka  \),
mixes together the Fourier modes of both \( \varphi  \) and \( \psi  \) 
\begin{eqnarray*}
\!\!\!E^{\text {int}}(\Delta ,\gamma )[\varphi ]\!\! &\! = \!& \!\!\frac{-\epsilon ^{2}}{2(1-\alpha )\ell ^{2D}}
\!\sum _{\ka \neq 0} (R^{T}_{\ka }\pd P_{\ka }\pd R_{\ka }+I^{T}_{\ka }\pd P_{\ka }\pd I_{\ka })\\
R_{\ka } & = & \left[ \Re \left( \widetilde{\varphi }_{\ka }\right);  
\Re \left( \widetilde{\psi }_{\ka }\right) \right] ^{T}\\
I_{\ka } & = & \left[ \Im \left( \widetilde{\varphi }_{\ka }\right); 
\Im \left( \widetilde{\psi }_{\ka }\right) \right] ^{T}\\
P_{\ka } & = & \left[ \begin{array}{cc}
L_{\ka } & M_{\ka }\\
M_{\ka } & N_{\ka }
\end{array}\right]   
\end{eqnarray*}
 where \( \Re  \) and \( \Im  \) represent the real and imaginary part of
a complex number. The components \( L_{\ka } \), \( M_{\ka } \), \( N_{\ka } \)
depend both on the applied-strain parameters (maximum or actual ones), and the
wave-vector \( \ka  \).   In anticipation of Paper III, 
it is convenient to introduce   
\[
\omega =\frac{\kappa _{3}}{\kappa_1}\frac{\gamma }{\Delta } \]
as the shear-strain variable and to define the parameter 
$$
c= \kappa_2/\kappa_1 = 1 + \frac{2\mu (\lambda + \mu) (\lambda _+ 2 \mu)}{\lambda^3}$$
where $c>1$.  The components of the matrix $P_{\ka}$ are then 
\begin{eqnarray*}
&&L_{\ka }  (\Delta ,\omega )=  \Delta ^{2}\kappa ^{2}_{1}(1-\alpha u^{2}_{\ka })\\
&&M_{\ka }  (\Delta ,\omega )=  \Delta ^{2}\kappa _{1}^2 
u_{\ka }\left[ (1-\alpha )c -\alpha v_{\ka }\omega \right] \\
&&N_{\ka }  (\Delta ,\omega )=  \Delta ^{2}\kappa_1^2 
\left[ (1-\alpha )c^2 \right.\\
&& \qquad \qquad \qquad \qquad \left.  +2(1-\alpha ) c 
 v_{\ka }\omega+
  (1-\alpha v^{2}_{\ka })\omega ^{2}\right] 
\end{eqnarray*}
where \( u_{\ka }=\cos (2\theta _{\ka }) \) and  \( v_{\ka }=\sin (2\theta _{\ka }) \)
are functions that characterize the orientation of \( \ka  \) through its polar
angle \( \theta _{\ka } \).  

Note that all terms contributing to the Hamiltonian have been written
in a dimensionless form in which energy density \( E_{j} \), like stress, is
measured in units of \( \lambda +2\mu  \).

\appendix
\section{Effective moduli of a cracked cell\label{ssec:Modulii}}

A crack is modeled here as an elongated ellipse having a major axis of length 
$d$ and a minor axes of length 
\( w \) in the limit that \( w/d\ll 1 \) which corresponds
to a so-called ``penny-shaped'' crack. Its long axis is by convention oriented
along \( \hat{\eu }_{1} \) if locally \( \varphi >0 \), and along \( \hat{\eu }_{2} \)
if \( \varphi <0 \). The unit cell is a square whose sides are colinear with
\( (\hat{\eu }_{1};\hat{\eu }_{2}) \), and has a size \( \Lambda \gg d \)
since the crack is taken to be small. The interior of the crack is supposed
to be much more compliant than the embedding matrix and all plastic deformation
will be ignored; \emph{i.e.}, there is no residual stress or strain allowed
for in the cracked system when it is unloaded to zero applied stress.

Denoting as usual the volume average of a quantity with an overbar, we seek
to determine the elastic-stiffness tensor \( \ce  \) of a cell as defined 
through  the relation 
\begin{equation}
\label{eq:def,modules,2}
\overline{\ta }_{ij}=\ce _{ijkl}\overline{\ep }_{kl}.
\end{equation}
 The region inside the crack is  occupied by a uniform 
material of stiffness \( \ce ^{1} \) while the intact matrix surrounding
the crack  is occupied by a material of 
stiffness \( \ce ^{1} \).  Upon denoting $v$ the volume fraction of the 
crack in the cell, we obtain directly  
\begin{equation}
\label{eq:tau,moyen,eps,ssvolumes}
\overline{\ta }_{ij}=(1-v)\ce _{ijkl}^{0}\overline{\ep }^{0}_{kl}+v\ce _{ijkl}^{1}\overline{\ep }^{1}_{kl}.
\end{equation}
  Eshelby
 \cite{Esh57} demonstrates that the strain \( \ep ^{1} \) inside an elliptic inclusion
is uniform while Wu \cite{Wu66}  relates this strain to the strain at infinity by a tensor \( \te  \) 
\begin{equation}
\label{eq:Wueq}
\overline{\ep }^{1}_{ij}=\te _{ijkl}\ep _{kl}^{\infty }.
\end{equation}
 With cracks considered as small inclusions in their embedding cell (\( v\ll 1 \)),
the approximation \( \ep ^{\infty }\simeq \overline{\ep } \) is valid to leading
order in the above, so that 
\begin{eqnarray}
\overline{\ep }^{0}_{ij} & = & \vv _{ijkl}\overline{\ep }_{kl}\label{eq:eps1,epsmoyen} \\
(1-v)\vv _{ijkl} & = & \left( \delta _{ik}\delta _{jl}-v\te _{ijkl}\right) \label{eq:def,tenseur,deux} 
\end{eqnarray}
 Using Eqs.\ (\ref{eq:Wueq}) and (\ref{eq:eps1,epsmoyen}) for the average
deformation in and out of the inclusion, Eq.\ (\ref{eq:tau,moyen,eps,ssvolumes})
has the desired linear form of Eq.\ (\ref{eq:def,modules,2}) with an effective
stiffness tensor given by 
\begin{eqnarray}
\ce _{ijkl} & = & \ce ^{0}_{ijkl}-v\left( \ce _{ijmn}^{0}-\ce _{ijmn}^{1}\right) \te _{mnkl}\nonumber \\
 & \simeq  & \ce ^{0}_{ijkl}-v\ce _{ijmn}^{0}\te _{mnkl}.\label{eq:modules,effectifs} 
\end{eqnarray}
 This approximation is justified under the hypothesis that the material inside
the inclusion (air) is far more compliant than the host material (solid silicate). These relations
are valid in any space dimension \( D \) . The two-dimensional case of interest
to us here can be obtained from the three-dimensional Wu-Eshelby results 
by working with a three-dimensional ellipsoidal inclusion having semi-axes of
linear dimension \( a=d/2 \); \( b=w/2 \); \( c=h/2 \) embedded within a
cell of dimension \( \Lambda \times \Lambda \times h \) in the limit that \( h\gg \Lambda  \).
In this limit, the three-dimensional problem becomes one in two dimensions.

 Wu expresses his tensor \( \te  \) in terms of a  tensor  \( \es  \)  defined by Eshelby 
\begin{eqnarray}
 &  & T_{ijij}=\frac{1}{2(1-2S_{ijij})}\label{eq:T1212}  \, \mbox{ when } \, \, i\neq j \\
 &  & \left( \begin{array}{ccc}
T_{1111} & T_{1122} & T_{1133}\\
T_{2211} & T_{2222} & T_{2233}\\
T_{3311} & T_{3322} & T_{3333}
\end{array}\right) =\nonumber \\
 &  & \qquad \left( \begin{array}{ccc}
1-S_{1111} & -S_{1122} & -S_{1133}\\
-S_{2211} & 1-S_{2222} & -S_{2233}\\
-S_{3311} & -S_{3322} & 1-S_{3333}
\end{array}\right) ^{-1}.  
\end{eqnarray}
The Eshelby \cite{Esh57} tensor components are defined 
\begin{eqnarray}
S_{1111} & = & Qa^{2}I_{aa}+RI_{a}\label{eq:defS1111} \\
S_{1122} & = & Qb^{2}I_{ab}-RI_{a}\label{eq:defS1122} \\
S_{1212} & = & \frac{Q}{2}(a^{2}+b^{2})I_{ab}+\frac{R}{2}(I_{a}+I_{b})\label{eq:defS1212}
\end{eqnarray}
 with similar expressions for the remaining components obtained through the
permutation of \( a,b,c \) and \( 1,2,3 \).  In the notation of the present 
paper, the various parameters of Eqs.\ (\ref{eq:defS1111})--(\ref{eq:defS1212}) 
 are defined 
\begin{equation}
\label{eq:defQ,R}
Q=\frac{3}{8\pi (1-\sigma _{p})}\mbox {and}\, R=\frac{1-2\sigma _{p}}{8\pi (1-\sigma _{p})}
\end{equation}
 where \( \sigma _{p}=\lambda /2(\lambda +\mu ) \) is the Poisson's ratio of
the solid material (assumed isotropic), and  
\begin{eqnarray}
I_{a} & = & 2\pi ab\int ^{\infty }_{0}\frac{du}{(a^{2}+u)D }\label{eq:defIa2D} \\
I_{aa} & = & 2\pi ab\int ^{\infty }_{0}\frac{du}{(a^{2}+u)^{2}D }\label{eq:defIaa2D} \\
I_{ab} & = & \frac{2}{3}\pi ab\int ^{\infty }_{0}\frac{du}{(a^{2}+u)(b^{2}+u)D }\label{eq:defIab2D} 
\end{eqnarray}
 with 
 $D =\sqrt{(a^{2}+u)(b^{2}+u)}$ and $
 I_{c}=I_{ac}=I_{bc}=I_{cc}=0$.  
 Similar expressions are obtained for \( I_{b} \) and \( I_{bb} \) by replacing
\( a \) and \( b \) in the above.
These elliptic integrals are evaluated to the leading  order in the small aspect
ratio \( \delta =b/a \) which gives 
\begin{eqnarray}
&&I_{a}  =  4\pi \delta, \mbox{\hskip3mm} 
I_{b}  =  4\pi (1-\delta )\label{eq:Ib2D} \\
&&I_{aa}  =  \frac{4\pi }{3a^{2}}2\delta, \mbox{\hskip3mm}  
I_{bb}  =  \frac{4\pi }{3b^{2}}, \mbox{\hskip3mm} 
 I_{ab}  =  \frac{4\pi }{3a^{2}}(1-2\delta ).\quad \, \, \, \label{eq:Iab2D} 
\end{eqnarray}
 Defining parameters \( q \) and
\( r \) by 
\begin{equation}
\label{eq:defq,r}
q=4\pi Q=\frac{3}{2(1-\sigma _{p})}\mbox {and}\, r=4\pi R=\frac{1-2\sigma _{p}}{2(1-\sigma _{p})},
\end{equation}
 we obtain that to the leading order in $1/\delta$  
\begin{eqnarray}
T_{1212} & = & \frac{3}{4q}\frac{1}{\delta }\label{eq:T12D} \\
T_{2211} & = & \frac{\frac{q}{3}-r}{r(1+\frac{q}{3}-r)}\frac{1}{\delta }\label{eq:T22D} \\
T_{2222} & = & \frac{1}{r(1+\frac{q}{3}-r)}\frac{1}{\delta }.\label{eq:T32D} 
\end{eqnarray}
 All remaining components of \( \te  \) are either \( O(1) \) and therefore
negligible, or are unimportant for the components of \( \ce  \) related to
directions \( 1 \) and \( 2 \).

To get finally the deviation \( \dc  \) of the effective elastic moduli of
the cracked cell through Eq.\ (\ref{eq:modules,effectifs}), we note first that
\begin{equation}
\label{eq:v}
v=\frac{4\pi }{3}\frac{abc}{\Lambda ^{2}h}=\frac{2\pi }{3}\frac{a^{2}}{\Lambda ^{2}}\frac{b}{a}=\frac{\pi }{6}\frac{d^{2}}{\Lambda ^{2}}\delta =\frac{\pi }{6}\epsilon \psi \delta 
\end{equation}
 where we recall that \( d=2a \) is the crack's length, \( w=2b \) its width, 
and \( \delta  \) its aspect ratio. It is through this expression
that the small parameter \( \epsilon =(d_{m}/\Lambda )^{2}\ll 1 \) enters the
Born series. Note that \( \psi =\left| \varphi \right| =(d/d_{m})^{2} \) characterizes
the extent of the crack. The third dimension of \( h=2c \) goes to infinity
in order to obtain the two-dimensional limit of this three-dimensional system.

Replacing \( q \) and \( r \) by their expressions in terms of the Lam\'{e}
parameters \( \lambda ,\, \mu  \), and using by convention \( \lambda +2\mu  \)
as the stress unit,  the crack-induced perturbations of the cell moduli are   
\begin{eqnarray}
 &  & \dc _{2222}=-v\ce ^{0}_{2222}\te _{2222}=-\frac{\pi }{6}\epsilon \psi 
\frac{\lambda +2\mu }{\mu }\label{eq:dc2D1} \\
 &  & \dc _{1111}=-v\ce ^{0}_{1122}\te _{2211}=-\frac{\pi }{6}\epsilon \psi 
\frac{\lambda ^{2}}{\mu (\lambda +2\mu )}\label{eq:dc2D2} \\
 &  & \dc _{1122}=-v\ce ^{0}_{1122}\te _{2222}=-\frac{\pi }{6}\epsilon \psi 
\frac{\lambda }{\mu }\label{eq:dc2D3}\\ 
 &  & \dc _{2211}=-v\ce ^{0}_{2222}\te _{2211}=-\frac{\pi }{6}\epsilon \psi \frac{\lambda }{\mu }\label{eq:dc2D4} \\
 &  & \dc _{1212}=-v(\ce _{1212}^{0}\te _{1212}+\ce _{1221}^{0}\te _{2112})\nonumber \\
 &  & \qquad \, \, \, \, \, \, =-\frac{\pi }{6}\epsilon \psi \frac{1}{2}\frac{\mu }{\lambda +\mu }\label{eq:dc2D5} 
\end{eqnarray}
with  all other terms being zero except those obtained by the necessary symmetries
under exchange of the two first or two last indexes.

Using the dimensionless constants \( \alpha  \) defined in Eq.\ (\ref{eq:def,alpha})
and introducing the positive dimensionless coefficients \( \eta _{i} \)
\begin{eqnarray}
\eta _{1} & = & \frac{\pi }{6}\frac{\lambda ^{3}+\mu (\lambda +2\mu )^{2}}{\lambda \mu (\lambda +2\mu )}\nonumber \\
\eta _{2} & = & \frac{\pi }{6}\frac{\lambda +2\mu }{\lambda }\nonumber \\
\eta _{3} & = & \frac{\pi }{12}\frac{\lambda +2\mu }{\lambda +\mu },
\end{eqnarray}
 we obtain at last the deviation of the elastic moduli of a cell containing
a crack with long axis oriented along \( \hat{\eu }_{1} \) (corresponding to
a positive \( \varphi  \)) 
\begin{equation}
\label{eq:dc,phi=1}
\dc =\left( \begin{array}{ccc}
\eta _{2}-2\eta _{1} & -(2\alpha -1)\eta _{2} & 0\\
-(2\alpha -1)\eta _{2} & -\eta _{2} & 0\\
0 & 0 & -(1-\alpha )\eta _{3}
\end{array}\right) \epsilon \psi 
\end{equation}
 The expression for both possible orientations of the cracks is straightforward.
Orienting the crack along \( \hat{\eu }_{2} \) instead \( \hat{\eu }_{1} \)
is equivalent to exchanging the 1 and 2 indices in the components of \( \dc  \)
which results in an exchange of the components \( \dc _{1111} \) and \( \dc _{2222} \),
all remaining components of \( \dc  \) being unaffected by this change. Separating
both expressions of \( \dc  \) into symmetric and antisymmetric parts, and
noting that \( \dc =0 \) trivially when \( \varphi =0 \) (no crack), we obtain
the general expression used in Eqs.\ (\ref{eq:C,A,B,phi})--(\ref{eq:B,simpl}).

\bibliographystyle{apsrev}

\end{document}